\documentclass{iopconfser}
\usepackage{units}
\usepackage{amsmath}
\usepackage{graphicx}
\usepackage[colorlinks=true,allcolors=black]{hyperref}

\begin{document}

\title{High-accuracy multi-ion spectroscopy with mixed-species Coulomb crystals}

\author{J.~Keller$^{1}$, H.~N.~Hausser$^{1}$, I.~M.~Richter$^{1}$, T.~Nordmann$^{1}$, N.~M.~Bhatt$^{1}$, J.~Kiethe$^{1}$, H.~Liu$^{1}$, E.~Benkler$^{1}$, B.~Lipphardt$^{1}$, S.~D\"orscher$^{1}$, K.~Stahl$^{1}$, J.~Klose$^{1}$, C.~Lisdat$^{1}$, M.~Filzinger$^{1}$, N.~Huntemann$^{1}$, E.~Peik$^{1}$, and T.~E.~Mehlst\"aubler$^{1,2,3}$}

\affil{$^1$Physikalisch-Technische Bundesanstalt, Bundesallee 100, 38116 Braunschweig, Germany}
\affil{$^2$Institut f\"ur Quantenoptik, Leibniz Universit\"at Hannover, Welfengarten 1, 30167 Hannover, Germany}
\affil{$^3$Laboratory of Nano and Quantum Engineering, Leibniz Universtit\"at Hannover, Schneiderberg 39, 30167 Hannover, Germany}

\email{jonas.keller@ptb.de, tanja.mehlstaeubler@ptb.de}

\begin{abstract}
  Multi-ion optical clocks offer the possibility of overcoming the low signal-to-noise ratio of single-ion clocks, while still providing low systematic uncertainties. We present simultaneous spectroscopy of up to four ${}^{115}$In${}^+$ clock ions in a linear Coulomb crystal, sympathetically cooled with ${}^{172}$Yb${}^+$ ions. In first clock comparisons, we see agreement below $1\times10^{-17}$ with results obtained using a single In${}^+$ ion, for which we have evaluated the systematic uncertainty to be $2.5\times10^{-18}$. Operation with four clock ions reduces the instability from $1.6\times10^{-15}/\sqrt{t/(\unit[1]{s})}$ to $9.2\times10^{-16}/\sqrt{t/(\unit[1]{s})}$. We derive a model for decay-related dead time during state preparation, which matches the observed scaling of instability with clock ion number $N$, and indicates that $1/\sqrt{N}$ scaling can be achieved with the addition of a repump laser.
\end{abstract}

\section{Introduction}
Coulomb crystals provide a scalable approach for storing multiple clock ions in a well-controlled environment for interrogation with low systematic uncertainties. They also straightforwardly enable coupling to ions of a different species for sympathetic cooling, quantum logic operations or in-situ measurements of environmental conditions. Clocks based on Coulomb crystals have gained interest in recent years \cite{Arnold2015, Aharon2019, Cui2022, Leibrandt2024, Steinel2023}, as the use of multiple clock ions reduces quantum projection noise (QPN) \cite{Itano1993}, the limiting instability contribution of current single ion clocks. A given level of statistical uncertainty can thus be reached with shorter averaging times, which would otherwise need to be prohibitively long to match the decreasing systematic uncertainties of these systems. Due to its low sensitivities to electric field gradients, In${}^+$ has been identified as one of the suitable candidates for a high-accuracy clock based on Coulomb crystals \cite{Herschbach2012}, with potential frequency uncertainties in the low $10^{-19}$ range \cite{Keller2019}. Here, we report on spectroscopy and clock operation with multiple In${}^+$ ions, sympathetically cooled with Yb$^+$ ions. We discuss site-resolved clock spectroscopy signals in section \ref{sec:spectroscopy}. In section \ref{sec:clock_operation}, we present initial clock operation with multiple In${}^+$ ions, compared against a Sr lattice and an Yb$^+$ ion clock. We investigate the scaling of the instability with clock ion number, derive a model which describes the current limitations due to dead time, and discuss our plans to overcome this limit.

\section{\label{sec:spectroscopy}Spectroscopy of multiple clock ions in a mixed-species chain}
\begin{figure}
  \centerline{\includegraphics[width=\textwidth]{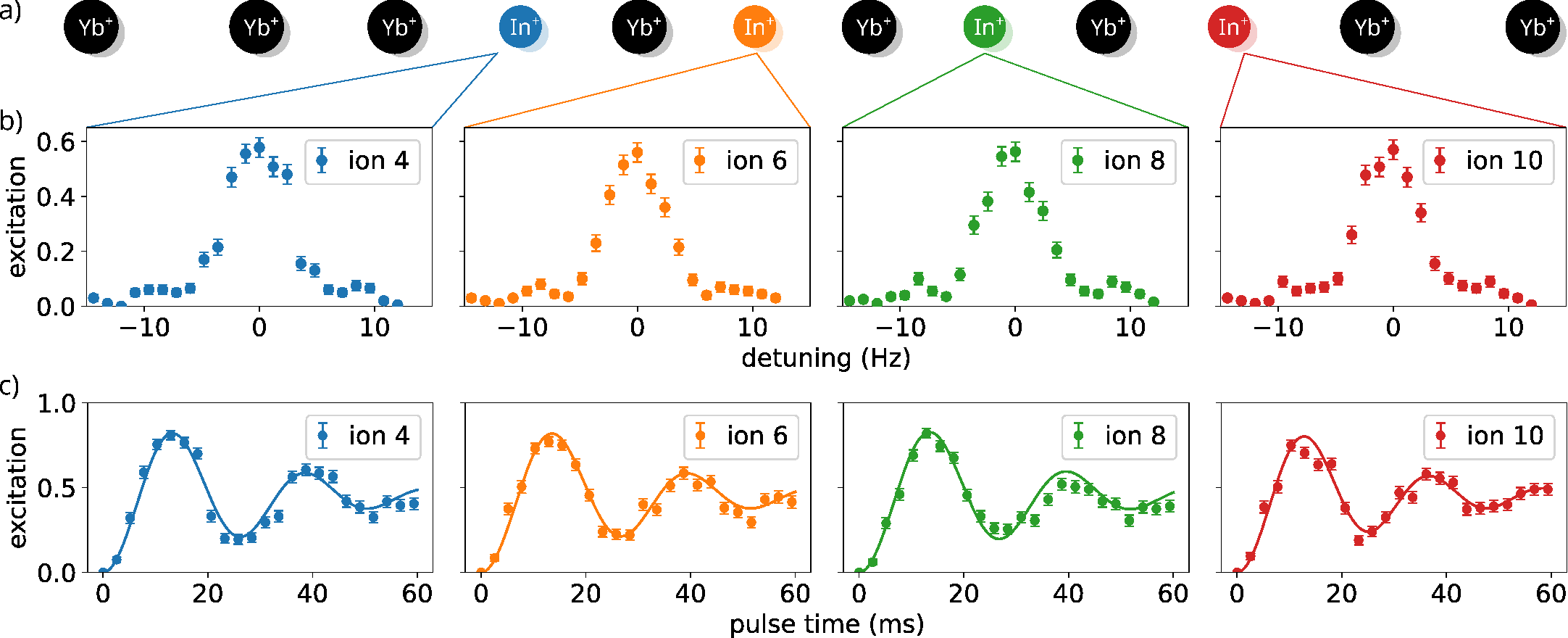}}
  \caption{\label{fig:scan_flops}Spectroscopy with four In${}^+$ ions. a) Mixed-species crystal consisting of four In${}^+$ clock ions and eight Yb${}^+$ cooling ions. b) Site-resolved frequency scan of the clock transition ($m_F=m_{F^\prime}=+9/2$ component). The frequency axis offset is the same for all plots. c) Site-resolved Rabi oscillations compared to the theoretical expectation at $1.3$ times the Doppler temperature for all $2N=24$ radial modes, scaled by $0.9$ to match the observed contrast. Error bars depict the quantum projection noise.}
\end{figure}

We operated the clock with three different crystal compositions, containing one, two, and four In${}^+$ clock ions, respectively. All experiments use the spectroscopy sequence described in \cite{Hausser2025}, which allows independent evaluation of the individual ion signals and ensures a reproducible permutation of clock and cooling ion positions within the crystal. Denoting ${}^{115}$In${}^+$ ions with ``$0$'' and ${}^{172}$Yb${}^+$ with ``$1$'', we use the permutations $1011$, $101011$ and $111010101011$ (or their mirror images, which are equivalent in terms of all relevant characteristics). These provide sufficient participation of the sympathetic cooling ions in all modes of motion. We apply Doppler cooling for $\unit[50]{ms}$ duration on the ${}^2S_{1/2}\leftrightarrow {}^2P_{1/2}$ transition in Yb${}^+$, with a power corresponding to the saturation intensity $I_\mathrm{sat}$ in the first half, followed by a linear ramp down to $0.1\times I_\mathrm{sat}$. Most of the cooling light is applied along a radial direction, with projections onto both radial trap principal axes. An additional beam at $0.3$ times the intensity provides cooling of the axial motion. The cooling is followed by interrogation of the clock transition. The clock laser addresses the crystal globally from the same direction as the radial cooling beam and has a waist of $w_z=\unit[820]{\mu m}$ along the trap axis. Figure \ref{fig:scan_flops} shows measurements with the 4In${}^+$-8Yb${}^+$ crystal, which is confined with secular frequencies (for the ${}^{172}$Yb${}^+$ center-of-mass modes) of $(\omega_\mathrm{ax}, \omega_\mathrm{rad1}, \omega_\mathrm{rad2})/2\pi=$ $(\unit[110]{kHz}, \unit[821]{kHz}, \unit[833]{kHz})$ and has an extent of $\unit[77]{\mu m}$. Note that the ion numbering is reversed for iterations which use the mirrored permutation. Figure \ref{fig:scan_flops}b shows a frequency scan over the $m_F=m_{F^\prime}=+9/2$ Zeeman component of the ${}^1S_0\leftrightarrow {}^3P_0$ clock transition, using $\unit[150]{ms}$ square pulses. All clock ions contribute with a contrast of ca.~$0.6$, which matches that seen in operation with a single clock ion \cite{Hausser2025}.

Figure \ref{fig:scan_flops}c shows Rabi oscillations on resonance of the same transition. The solid lines are added for a coarse temperature estimate: They use a model which takes into account spontaneous decay and thermal dephasing due to the $2N=24$ radial modes of motion onto which the clock laser wave vector has nonzero projection (Debye-Waller effect \cite{Wineland1998}), assuming the set of motional modes along each radial principal axis to be at $1.3$ times the respective Doppler temperature. The overall contrast of the model was reduced by a factor of $0.9$ to fit the data, which could be caused by imperfect state preparation. For the parameters used in the frequency scans of Fig.~\ref{fig:scan_flops}b, the model reproduces the observed contrast of $0.6$. The agreement of model and data would be further improved by more precise knowledge of the individual mode temperature degrees of freedom, e.g., through thermometry.

For the 1In${}^+$-3Yb${}^+$ crystal, we have evaluated the systematic uncertainty to be $2.5\times10^{-18}$ \cite{Hausser2025}. While a similar analysis has not yet been performed for the larger crystals, we expect most significant contributions to be independent of the ion number, with the exception of the thermal time dilation shift. The agreement with the Rabi oscillation model assuming temperatures close to the Doppler limit suggests that shifts are at the low $10^{-18}$ level, but more detailed analysis of the motional excitations is required to reduce the corresponding uncertainties. Lower temperatures of the relevant modes can be achieved with direct cooling of In${}^+$ on the ${}^1S_0\leftrightarrow{}^3P_1$ transition \cite{Peik1999} in the intermediate cooling regime (transition linewidth on the order of the secular frequencies) \cite{Kulosa2023}, which is expected to reduce the time dilation shift to the low $10^{-19}$ range \cite{Keller2019}.

\section{\label{sec:clock_operation}Multi-ion clock operation}
\begin{figure}
  \centerline{\includegraphics[width=\textwidth]{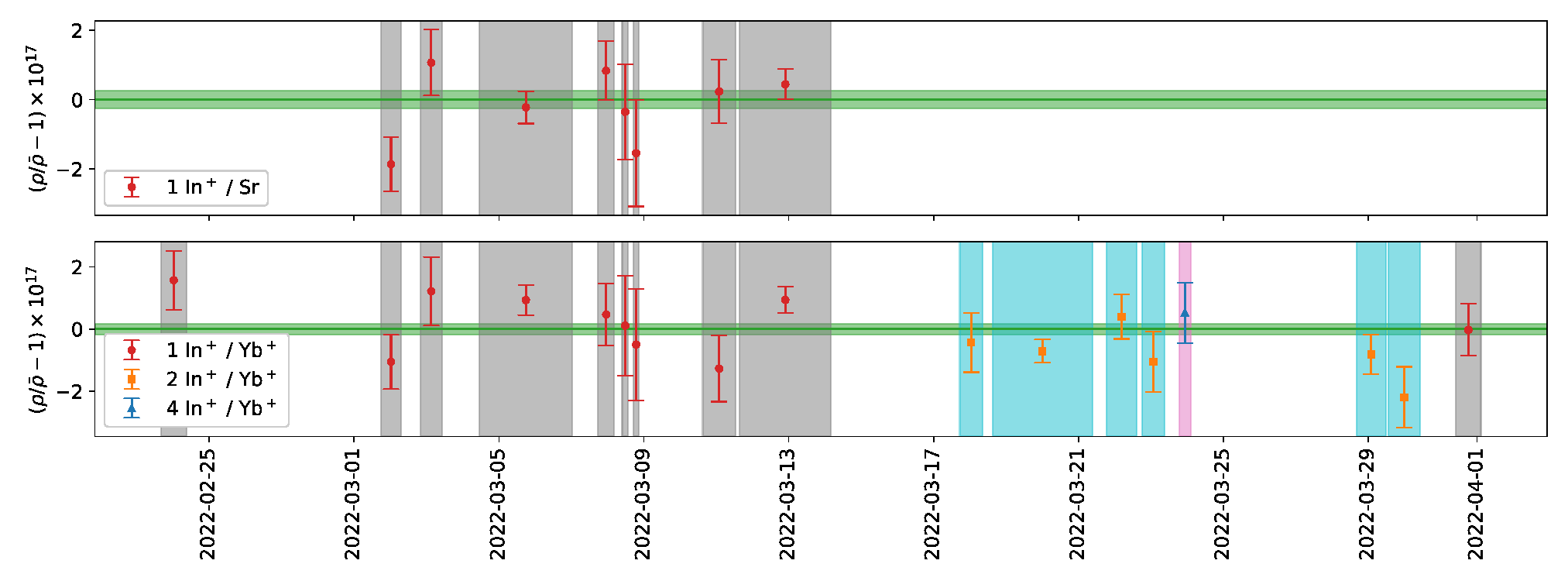}}
  \caption{\label{fig:ratios} Optical frequency comparisons to ${}^{87}$Sr (top) and ${}^{171}$Yb${}^+$ E3 (bottom) using different numbers of clock ions. Vertical shaded areas show the durations of the respective datasets. The green lines and shaded areas depict the weighted means and corresponding uncertainties. All shown uncertainties are purely statistical, based on the observed white frequency noise.}
\end{figure}

The crystal compositions with one, two, and four In${}^+$ ions have been used in optical clock operation. Figure \ref{fig:ratios} shows first measurements in which the In${}^+$ clock is compared to a ${}^{87}$Sr lattice clock \cite{Schwarz2022} and an ${}^{171}$Yb${}^+$ single-ion clock based on the electric octupole (E3) transition \cite{Sanner2019}. Data before March 14$^{\mathrm{th}}$, 2022 are those used in the evaluation of \cite{Hausser2025}. The average In${}^+$ / Yb${}^+$ frequency ratio observed in the measurements with multiple In${}^+$ ions agrees with the single-ion result to within less than $1\times10^{-17}$. Agreement at the $10^{-18}$ level will be tested in future campaigns. In the remainder, we focus instead on the observed multi-ion clock instabilities.

\subsection{Decay-related dead times in state preparation}
After successful excitation attempts, the ions need to be re-initialized in the ground state. The lifetime of $\tau=\unit[195]{ms}$ \cite{Becker2001} is too short for an alternative approach which inverts the role of the two clock states in the subsequent iteration. Optical repumping can be used for deterministic initialization, but was not available during the time of the measurements presented here. Alternatively, the electronic excitation can be transferred to the state of motion and removed via laser cooling \cite{Chou2017}, but this requires additional experimental overhead and ground-state cooling. In the initial In$^+$ multi-ion clock operation, state preparation was instead achieved by repeatedly monitoring the state of the clock ions while waiting for natural decay. In the following, we investigate theoretically and experimentally the frequency instability implications of the associated dead time.

\subsection{Average decay times for subsets of a clock ion ensemble}
Site-resolved detection allows clock operation with a subset of $n_\mathrm{min}$ out of $N$ clock ions, which can yield a lower instability than waiting for all clock ions to decay before every cycle. We therefore first derive the average delay until $n_\mathrm{min}$ decays have occurred.\\

\begin{figure}
  \centerline{\includegraphics[width=\textwidth]{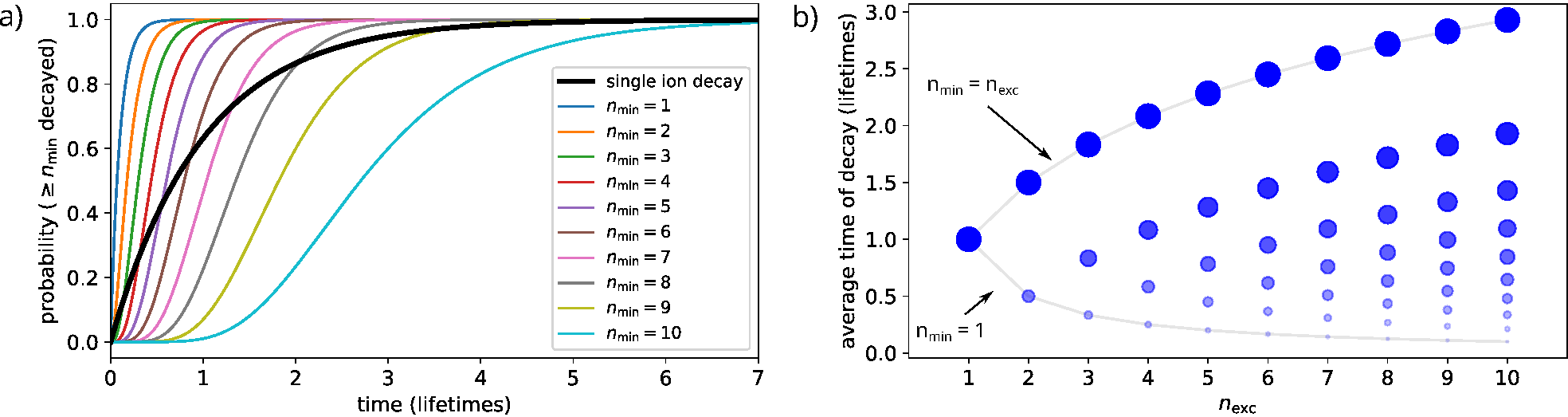}}
  \caption{\label{fig:decay_times} Decay times in ensembles of ions. a) Temporal evolution of the probability for at least $n_\mathrm{min}$ out of $n_\mathrm{exc}=10$ ions to have decayed (see Eq.~(\ref{eq:luh_decay_at_least_n})). b) Average time at which the $n_\mathrm{min}$th ion decays (see Eq.~(\ref{eq:luh_average_decay_time})) for $n_\mathrm{exc}=1\ldots10$ and $n_\mathrm{min}=1\ldots n_\mathrm{exc}$. The symbol sizes correspond to the respective fractions $n_\mathrm{min}/n_\mathrm{exc}$.}
\end{figure}

For a two-level atom which is in the excited state at $t=0$, the excited state population follows $P_{e,1}(t)=e^{-t/\tau}=1-P_{g,1}(t)$. The probability to find $n$ out of $n_\mathrm{exc}$ initially excited atoms in the ground state is thus
\begin{align}
  P_g(n, n_\mathrm{exc}, t)&=\binom{n_\mathrm{exc}}{n}P_{e,1}^{n_\mathrm{exc}-n}(t)P_{g,1}^n(t)\\
  &=\binom{n_\mathrm{exc}}{n}e^{-(n_\mathrm{exc}-n)t/\tau}\left(1-e^{-t/\tau}\right)^n\\
  &=\sum_{i=0}^n\binom{n_\mathrm{exc}}{n}\binom{n}{i}\left(-1\right)^ie^{-(n_\mathrm{exc}-n+i)t/\tau}\;\textnormal{,}
\end{align}
and the probability for \emph{at least} $n_\mathrm{min}$ atoms to have decayed is
\begin{equation}
  P_{g,\geq}(n_\mathrm{min}, n_\mathrm{exc}, t) = \sum_{j=n_\mathrm{min}}^{n_\mathrm{exc}}P_g(j, n_\mathrm{exc}, t) = \sum_{j=n_\mathrm{min}}^{n_\mathrm{exc}}\sum_{i=0}^j\binom{n_\mathrm{exc}}{j}\binom{j}{i}\left(-1\right)^ie^{-(n_\mathrm{exc}-j+i)t/\tau}\;\textnormal{,}
  \label{eq:luh_decay_at_least_n}
\end{equation}
as illustrated in Fig.~\ref{fig:decay_times}a. The average time at which the $n_\mathrm{min}$th decay occurs is given by
\begin{equation}
  T(n_\mathrm{min}, n_\mathrm{exc}) = \int_0^\infty t\left(\frac{\partial}{\partial t} P_{g,\geq}(n_\mathrm{min}, n_\mathrm{exc}, t)\right)dt = \tau\cdot\left(\sum_{j=n_\mathrm{min}}^{n_\mathrm{exc}}\;\sum_{\substack{i=0\;(j<n_\mathrm{exc})\\i=1\;(j=n_\mathrm{exc})}}^j\binom{n_\mathrm{exc}}{j}\binom{j}{i}\frac{\left(-1\right)^{(i+1)}}{n_\mathrm{exc}-j+i}\right)\;\textnormal{.}
  \label{eq:luh_average_decay_time}
\end{equation}

Figure \ref{fig:decay_times}b shows the average decay times (\ref{eq:luh_average_decay_time}) for $n_\mathrm{exc}=1\ldots10$ and $n_\mathrm{min}=1\ldots n_\mathrm{exc}$.
In a final step, we consider that each clock probe on average only excites a fraction $n_\mathrm{exc}=p_\mathrm{exc}N$ of the ions. By averaging over the distribution of excited ion numbers, we obtain the average dead time during clock operation:
\begin{equation}
  T_\mathrm{decay}(n_\mathrm{min},N,p_\mathrm{exc}) = \sum_{n_\mathrm{exc}=0}^N P_\mathrm{exc}\left(n_\mathrm{exc}, N\right)\;T\left(n_\mathrm{min}-\left(N-n_\mathrm{exc}\right), n_\mathrm{exc}\right)\;\textnormal{,}
  \label{eq:luh_decay_dead_time}
\end{equation}
where $n_\mathrm{min}$ has been reduced by the amount $N-n_\mathrm{exc}$ of ions which were never excited, and we set $T(n_\mathrm{min}, N)=0$ for $n_\mathrm{min}\leq0$ (i.e.,~when the remaining number of ions ground state is enough to immediately continue with the next iteration). $P_\mathrm{exc}$ in (\ref{eq:luh_decay_dead_time}) is another binomial distribution:
\begin{equation}
  P_\mathrm{exc}(n_\mathrm{exc}, N)=\binom{N}{n_\mathrm{exc}}p_\mathrm{exc}^{n_\mathrm{exc}}\left(1-p_\mathrm{exc}\right)^{\left(N-n_\mathrm{exc}\right)}\;\textnormal{.}
\end{equation}

\subsection{Implications for multi-ion clock instabilities}
\begin{figure}
  \centerline{\includegraphics[width=.8\textwidth]{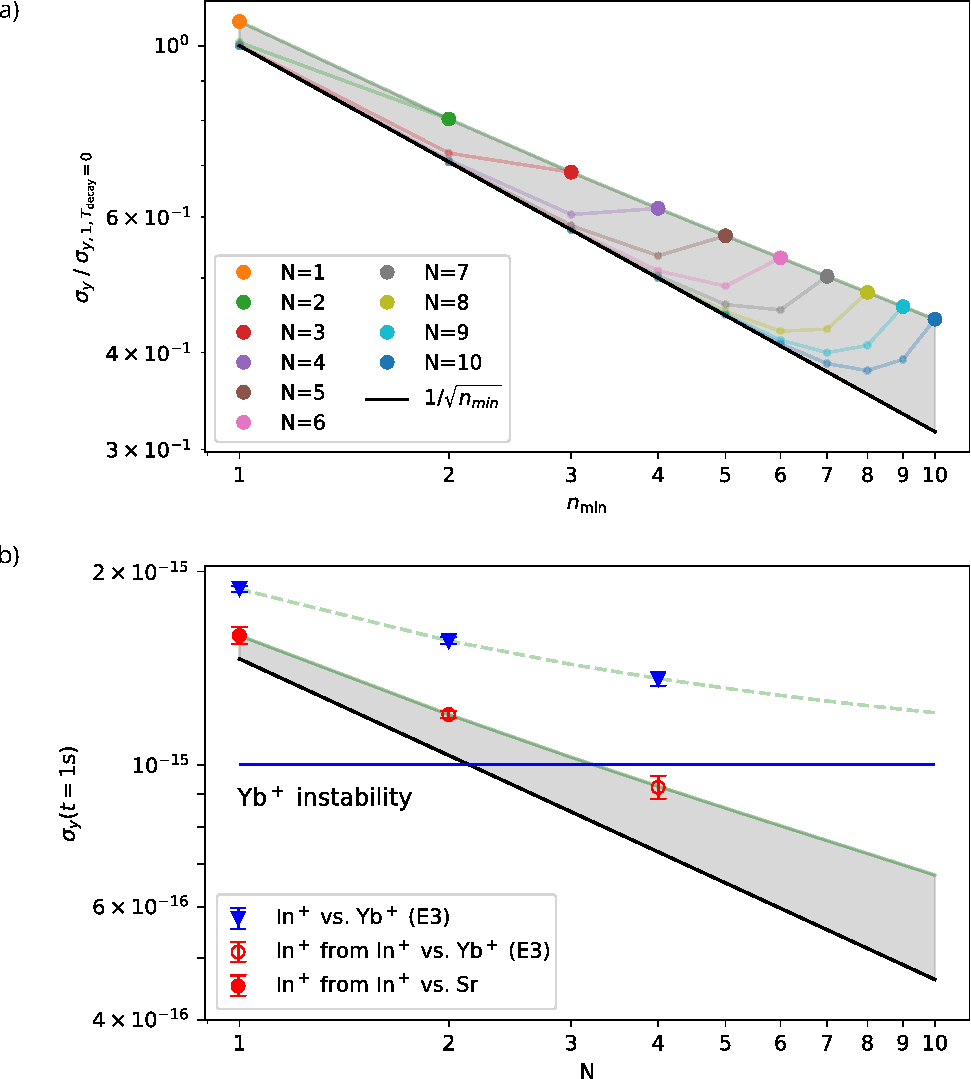}}
  \caption{\label{fig:multi_ion_instability} Multi-ion clock instabilities. a) Expected quantum projection noise limited instabilities in multi-ion clocks with atomic state initialization based on spontaneous decay (see Eq.~(\ref{eq:luh_decay_limited_instabilities})), assuming the experimental parameters of section \ref{sec:exp_instabilities}. Each clock sequence iteration pauses until at least $n_\mathrm{min}$ clock ions are in the ground state. The different colors depict manifolds with different total numbers of clock ions $N$. All values are normalized to the single-ion QPN instability under the assumption of instantaneous decay. The shaded area indicates the increase with respect to the $1/\sqrt{n_\mathrm{min}}$ scaling in the absence of this dead time (solid black line). Note that this treatment potentially overestimates the instabilities for $n_\mathrm{min}<N$ (small circles), but is accurate for $n_\mathrm{min}=N$ (large circles) (see text). b) Experimentally observed instabilities in multi-ion operation by comparison to a Sr lattice clock and an Yb$^+$ (E3) single-ion clock. The solid lines and shaded area correspond to those of a). Their normalization is derived from the observed instability with a single clock ion \cite{Hausser2025}. Solid lines refer to the instability of the In$^+$ clock, while the dashed line shows the joint instability of the In${}^+$ clock and Yb$^+$ single-ion clock. With the addition of a repump laser, the clock is expected to reach the instabilities of the solid black line.}
\end{figure}

While the increased signal of a larger ion ensemble reduces QPN, the additional dead time without deterministic clock state initialization reduces the rate of data acquisition and partially counteracts this advantage. The overall effect can be expressed as
\begin{equation}
  \sigma_y(t)=\frac{\sigma_{y,1}(t)}{\sqrt{n_\mathrm{min}}}\sqrt{\frac{T_c + T_\mathrm{decay}(n_\mathrm{min}, N, p_\mathrm{exc})}{T_c}}\;\textnormal{,}
  \label{eq:luh_decay_limited_instabilities}
\end{equation}
where $\sigma_{y,1}(t)\propto t^{-1/2}$ is the single-ion instability due to QPN at averaging time $t$, and $T_c$ is the cycle time in the absence of the decay-related dead-time. Figure \ref{fig:multi_ion_instability}a illustrates the trade-off when choosing $n_\mathrm{min}$ for a given total number of ions for our parameters (see \ref{sec:exp_instabilities}). The solid black line depicts the $1/\sqrt{n_\mathrm{min}}$ scaling expected without additional dead time. As their number increases, it can be beneficial to operate with fewer than the total number of clock ions as long as state initialization depends on spontaneous decay. \footnote{Note that this treatment does not include cases in which more than $n_\mathrm{min}$ ions are available occasionally and the fact that ions on ``discarded'' sites might become available before the next cycle due to decay or interaction with the globally applied clock laser. The instability values for $n_\mathrm{min}<N$ therefore constitute upper bounds, but neither of these effects occur in the case of $n_\mathrm{min}=N$ (solid green line) tested below.}

\subsection{\label{sec:exp_instabilities}Experimental results}
In clock operation with a single In$^+$ ion (cooled by three Yb$^+$ ions) and $t_p=\unit[150]{ms}$ probe pulses, we observe an average cycle time of $\unit[330]{ms}$ and an average excitation of $p_\mathrm{exc}=0.26$ at the probe detunings of $\delta=\pm1/(2t_p)$. Of the $\unit[180]{ms}$ dead time, $p_\mathrm{exc}\cdot\tau=\unit[51]{ms}$ can be attributed to decay. Most of the remaining dead time consists of Yb${}^+$ sympathetic cooling and In${}^+$ state detection, both of which can be subject to further optimizations.
Figure \ref{fig:multi_ion_instability}b shows the QPN-limited instability at $t=\unit[1]{s}$ averaging time of the $^{115}$In$^+$ clock as determined from those of the optical frequency ratio measurements of Fig.~\ref{fig:ratios}. For $N=1$, we infer $\sigma_y=1.6\times10^{-15}/\sqrt{t / (\unit[1]{s})}$ from the In${}^+$ / Sr ratio, using the Sr clock instability of $2.0\times10^{-16}\sqrt{t/(\unit[1]{s})}$ \cite{Schwarz2020} (filled red circle). The blue triangles show the instabilities of the In${}^+$ / Yb${}^+$ (E3) frequency ratio, to which the Yb${}^+$ single-ion clock contributes $1.0\times10^{-15}/\sqrt{t/(\unit[1]{s})}$. The In${}^+$ clock instabilities for $N=2$ and $N=4$ are derived from those data, since no comparisons to the lattice clock were available under these conditions (red open circles). With $N=4$, the instability is reduced to $\sigma_y=9.2(4)\times10^{-15}/\sqrt{t / (\unit[1]{s})}$. The solid lines and shaded area correspond to the theoretical expectations of Fig.~\ref{fig:multi_ion_instability}a, adjusted to the observed instability at $N=1$. The deviation from the ideal $\sigma_y\propto 1/\sqrt{N}$ scaling matches the expected behavior due to dead time increase. Thus, implementing a controlled clock state initialization and thereby reducing the dead time due to spontaneous decay, we expect to reach the instabilities depicted by the black line in Fig.~\ref{fig:multi_ion_instability}. 

\subsection{Conclusion}
In clock operation with multiple In${}^+$ ions, the observed instabilities match the limit set by state preparation through spontaneous decay, as described by our model. To overcome this limitation in future measurements, state preparation with a repump transition ($^3P_0\leftrightarrow{}^1P_1$) at $\unit[482]{nm}$ should be feasible \cite{Safronova} and will be investigated. Without any further technical improvements, this is expected to achieve the instabilities depicted by the solid black line of Fig.~\ref{fig:multi_ion_instability}b.

\section*{Acknowledgments}
We thank I. Vybornyi for helpful discussions. We acknowledge support by the projects 18SIB05 ROCIT and 20FUN01 TSCAC. These projects have received funding from the EMPIR programme co-financed by the Participating States and from the European Union’s Horizon 2020 research and innovation programme. We acknowledge funding by the Deutsche Forschungsgemeinschaft (DFG) under Germany’s Excellence Strategy – EXC-2123 QuantumFrontiers – 390837967 (RU B06) and through Grant No. CRC 1227 (DQ-mat, projects B02 and B03). This work has been supported by the Max-Planck-RIKEN-PTB-Center for Time, Constants and Fundamental Symmetries.

\clearpage
\bibliographystyle{iopart-num}
\providecommand{\newblock}{}

\end{document}